**Investigating Systematic Uncertainty and Experimental Design with Projectile Launchers**

Authors: Chad Orzel, Gary Reich, Jonathan Marr

Union College Department of Physics and Astronomy, Schenectady, NY 12308

The proper choice of a measurement technique that minimizes systematic and random uncertainty is an essential part of experimental physics. These issues are difficult to teach in the introductory laboratory, though: because most experiments involve only a single measurement technique, students are often unable to make a clear distinction between random and systematic uncertainties, or to compare the uncertainties associated with different techniques. In this paper, we describe an experiment suitable for an introductory college level (or advanced high school) course that uses velocity measurements to clearly show students the effects of both random and systematic uncertainties.

Projectile motion experiments are a staple of introductory physics laboratories, and generally involve either measuring the velocity of a projectile using a single technique, such as video analysis (1-3) or timing the fall of some object (4-6). These experiments tend to focus on making an accurate measurement of a known quantity using a single technique, and achieve high accuracy even with rather simple techniques; for example, Ref. 6 reports a measurement of *g* to within 1% by aggregating timing measurements from an entire class.

In our experiment, students compare different techniques for measuring the launch speed of a plastic ball in a way that highlights the different systematic uncertainties affecting the two techniques. The comparison between methods also illustrates how different measurement techniques produce different statistical uncertainty in the final measurement, with results that many students find surprising.

Students measure the speed of a plastic ball launched vertically from a spring-loaded projectile launcher (Pasco 6801) in two different ways: by using a digital stopwatch to measure the time that the ball spends in the air, and by using a two-meter stick mounted next to the launcher to measure the maximum height reached by the ball. The maximum height measurements are made by eye, with students typically standing on a lab table to have their eyes at the appropriate level. Each student records their own measurements of ten heights and ten times, and calculates the average time, the average height, and the average launch speed. When asked to predict which method they expect will give the best measurement, students almost invariably choose the time-of-flight method, reflecting a widespread belief that a digital readout guarantees greater accuracy.

The launch speed $v_0$ in terms of the time of flight $t$ found using simple constant acceleration kinematics is:

$$v_0 = \frac{1}{2}gt - \frac{h}{t} \tag{1}$$

where $h$ is the height of the launch point above the floor (about 0.23 m, much less than the maximum height) and $g$ the acceleration due to gravity. The launch speed in terms of the maximum height $H$ is:

$$v_0 = \sqrt{2gH} \tag{2}$$

From these equations we see that, to leading order, the launch speed increases linearly with the measured time, but only as the square root of the measured height. The effect of uncertainty in the measured height is thus smaller than that of a similar fractional uncertainty in the time of flight. Contrary to student expectations, then, the maximum height method is likely to give better results than the time of flight method.

In fact, the maximum height method gives substantially better results, as the fractional uncertainty in the measured height is smaller than that of the measured time. For the Pasco launcher on the medium setting, typical values of the time of flight are ~1.4 s with an uncertainty of 0.04 s, while typical heights are ~2.15 m with an uncertainty of 0.007 m (quoted uncertainties are the standard deviation of the mean of ten measurements). The corresponding velocity uncertainties for the time-of-flight and maximum height methods are ~0.2 m/s and ~0.01 m/s, respectively.

The difference in the statistical uncertainty of the two methods can be verified by direct inspection of the velocities calculated from the individual time and height measurements. The range of velocities calculated from even the best set of student time-of-flight measurements is much broader than the range of velocities calculated from maximum height measurements. For more mathematically advanced classes, students can use uncertainty propagation formulae to find the uncertainty in the velocity, and verify that the calculated uncertainty agrees with the standard deviation of the individual velocity measurements.

The addition of a third measurement allows us to demonstrate the effect of systematic timing errors. Each lab group of three students records one launch of the ball using a high-speed camera (Kodak Motion Corder/ Analyzer, SR series) recording at 250 frames per second. By measuring the time between the ball leaving the launcher and hitting the ground, students obtain the same time-of-flight value measured with the watch, without the systematic effects introduced by human reaction time (similar results could be obtained using a cheaper camera with a lower frame rate, and averaging several trials).

Figure 1 shows the launch speeds determined from data collected by 26 students in our algebra-based "Physics for Life Sciences" class. Filled circles represent the speed determined from the time of flight method using a stopwatch, open circles the speed determined from the maximum height method, and triangles the speed determined from the time of flight method using the high-speed camera. Each lab group of 2-3 students made a single measurement with the high-speed camera, which is plotted for all members of the group; the uncertainty assigned to the single camera measurement is based on the frame rate of the camera. The speeds measured by the maximum height method and time of flight using the high-speed camera are consistently close together, with differences between groups due in large part to the different launchers used. The speeds measured from the time of flight using the stopwatch have both larger uncertainties (due to the larger fractional uncertainty in timing) and large systematic offsets caused by differences in individual reaction times.

Taken together, these data provide a dramatic graphical demonstration of the difference between random and systematic errors, and also a demonstration of the importance of measurement technique. While students consistently expect the "low-tech" maximum height method to produce the largest uncertainty, it is in fact the best of the three techniques—only 2 of the 26 students whose height data were used for Figure 1 had uncertainties greater than the 0.02 m/s uncertainty for the high-speed camera measurements.

This experiment provides an excellent opportunity both to reinforce key concepts from kinematics and also to introduce important ideas in data reduction and uncertainty analysis. In the course of the lab, we introduce the idea of the standard deviation and standard deviation of the mean as ways of characterizing the statistical uncertainty in a set of measurements, and at the conclusion of the lab we show the class a graph like Figure 1, which serves as the basis for a discussion of systematic errors and experimental design. When time permits, we occasionally have students use their measured launch speeds to calculate the range of a projectile fired at some angle, and use their result to place a "target" to be hit with their launcher, which helps confirm the accuracy of the maximum height method.

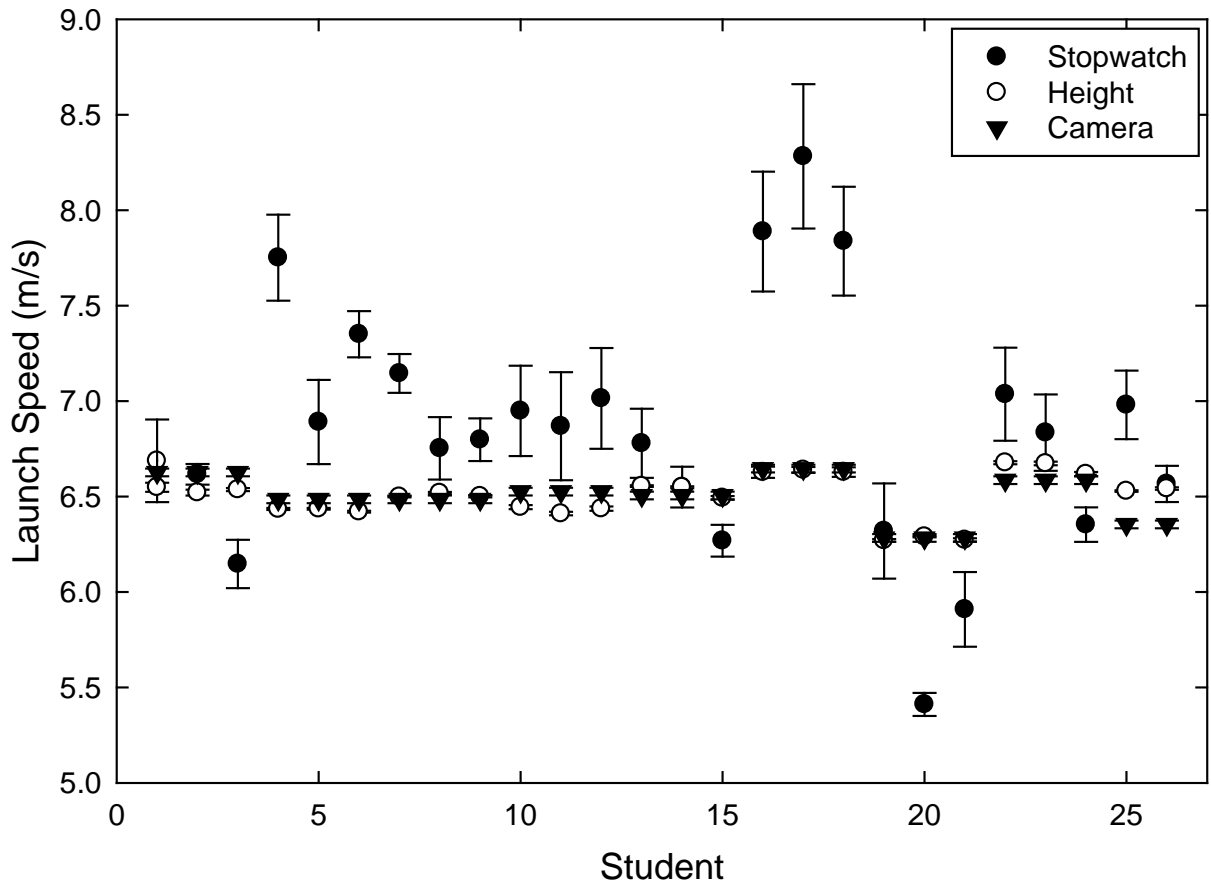

Figure 1: Launch speeds determined from data acquired by 26 students in the algebra-based "Physics for Life Sciences" class. Filled circles represent the speed determined from the time of flight measured with a stopwatch, open circles the speed determined from the maximum height method, and triangles the data determined from the time of flight method using a high-speed video camera. Error bars are 1-σ statistical uncertainties determined from the standard deviation of the mean of ten individual measurements. Each group of three students made a single measurement with the high-speed camera. Five different launchers of the same type were used to obtain these data, which accounts for some of the variation between groups.